# Persistent disruption of interspecific competition after ultra-low esfenvalerate exposure

Florian Schunck †, §, *; Matthias Liess †, §,

† Helmholtz Centre for Environmental Research (UFZ), Dept. of System-Ecotoxicology, Permoserstraße 15, 04318 Leipzig, Germany

§ Rheinisch-Westfälische Technische Hochschule (RWTH), Institute of Ecology & Computational Life Science, Templergraben 55, 52056 Aachen, Germany

## Abstract

Field and mesocosm studies repeatedly show that higher tier process reduce the predictive accuracy of toxicity evaluation and consequently their value for pesticide risk assessment. Therefore, understanding the influence of ecological complexity on toxicant effects is crucial to improve realism of aquatic risk assessment. Here we investigate the influence of repeated exposure to ecologically realistic concentrations of esfenvalerate on the similarly sensitive species *Daphnia magna* and *Culex pipiens* in a food limited and highly competitive environment. We show that significant perturbations in population development are only present close to the $EC_{50}$. In contrast, interspecific competition between species is already reduced at concentrations 3-4 orders of magnitude below the acute $EC_{50}$. We conclude that extremely low, environmentally relevant concentrations can disrupt species interactions. This toxicant mediated alteration of competitive balances in ecological communities may be the underlying mechanism for shifts in species distribution at ultra-low pesticide concentrations. A realistic risk assessment should therefore consider these processes in order to predict potential pesticide effects on the structure of communities.

## Introduction

Effect assessment is centered around screening for toxic effects with single species, single substance tests (European Food Safety Authority 2013). However, a growing number of studies indicate that whenever complex biological systems are exposed to a stressor, results diverge from expectations (Fleeger et al. 2003; Knillmann, Stampfli, Beketov, et al.

2012; Knillmann, Stampfli, Noskov, et al. 2012; Liess et al. 2013; Alexander et al. 2016; Arce-Funck et al. 2016; Vaugeois et al. 2020; Allen et al. 2021). Due to this gap between single species lab experiments and mesocosm experiments, inclusion of species interactions is one of the most important aspects of strengthening risk assessment (Gessner and Tlili 2016). Unfortunately, little progress has been made, as studies considering biological stressors such as species interactions are still underrepresented in literature (He et al. 2023).

The population state is an important co-variate for toxic effects of chemicals. Intraspecific competition can delay recovery of the population structure (Liess et al. 2006, Pieters and Liess 2006, Liess and Foit 2010). In the context of ecological communities, the competitive exclusion principle states that complete competitors (i.e., those that compete for exactly the same ecological niche) cannot coexist (Gause 1936; Hardin 1960). In natural ecosystems species diversify into their own niche, however, usually some overlap between shared resources remains, allowing for co-existence of competitors (MacArthur 1958; Hawlena et al. 2022). When such communities are exposed to toxicants, altered species–species interactions can therefore be expected. This is because usually one species will have a competitive advantage if exposed to a toxicant, due to differences in the species' sensitivity. Interspecific competition can delay recovery of species after disturbances (Knillmann, Stampfli, Noskov, et al. 2012) and increase toxic effects of pesticides (Knillmann, Stampfli, Beketov, et al. 2012). Under repeated lethal exposure to toxicants, the more sensitive species is gradually excluded, even when food density is abundant (Liess et al. 2013). In a synthetic freshwater community, the exposure to acute concentrations of an insecticide lead to reduced abundance in both competitors when they had a comparable sensitivity towards the toxicant and led to compensatory dynamics if sensitivities were different (Mano and Tanaka 2016).

How do pesticides alter interactions between competing species; do they cease or do they change when concentrations are far below levels that elicit acute effects? In the field, pesticide exposure 3 orders of magnitude below the $EC_{50}$ results in severe degradation of community composition with the loss of sensitive species (Liess et al. 2021). Recently, it has been shown that exposing *D. magna* populations at carrying capacity to esfenvalerate at concentrations 2 orders of magnitude below the acute $EC_{50}$ can lead to a long-term increases in population abundance (Schunck and Liess 2023). This reinforces the question of how sub-acute concentrations act at the ecological level of the community.

Our aim was to reveal the effects of ultra-low esfenvalerate concentrations on the population development of two competing species (*D. magna* and *C. pipiens*) in a food limited system with high competition between the species. For this we set up laboratory nanocosms and repeatedly exposed them with esfenvalerate concentrations as low as 3.5 orders of magnitude below the acute $EC_{50}$. The system state was monitored over a period of 4 months through non-invasive weekly monitoring of species abundance and measurement of physico-chemical parameters to assess the influence of environmental parameters on population development. Finally, effects of esfenvalerate on the interaction strength between the competing species *D. magna* and *C. pipiens* were investigated with Bayesian methods.

## Methods

**Experiment Design**

To study the effects of low doses of esfenvalerate on competing populations under limited availability of food and varying environmental conditions, 80 artificial 2–species systems were assembled in November 2020, under controlled temperature (20 ± 1 °C) and light conditions (16:8 day–night cycle). *Daphnia magna* and *Culex pipiens* were selected as competitors, both of which are common invertebrates that dwell in standing freshwater and brackish water bodies (Ebert 2022). While *D. magna* spends its entire life cycle in water, the species *C. pipiens* emerges after an approximately 20 day underwater larval stage as an adult mosquito and can reproduce without feeding on animal blood. Both species feed on suspended particles in the water column (Merritt et al. 1992; Ebert 2022) or moved close to the sediment to graze on organic particles of growing periphyton.

Each experimental unit consisted of a 5.5 L glass beaker (Harzkristall, Derenburg, Germany), filled with 1.5 kg of washed aquarium sand of 1–2 mm diameter. The sediment layer served as a habitat for microorganisms to facilitate self-purification of the systems as well as substrate for periphyton growth. Aachener Daphnien Medium (ADaM) (Klüttgen et al. 1994) was used as the test medium for the experiment. Throughout the duration of the experiment, the medium was not exchanged and kept at a constant volume of 3.5 L by replenishing the beakers with bi-distilled water on a weekly basis. The systems were covered with a polypropylene net to prevent escape of the adult mosquitoes. Two eyelets were embedded in the netting to grant access to the systems for measurement, sampling and supply of glucose solution. Additionally, a reaction vessel was fitted in the netting and immersed in the water column and filled with distilled water itself. This provided access for temperature monitoring without cross-contaminating the measurement device.

For 5 months, the systems were continuously colonized, while the systems were developing periphyton growth on the sediments, which served as a food source for the organisms. The systems were deliberately left to diverge from the initial homogeneous state to reflect random variation in environmental habitats. In contrast to previously conducted nanocosm experiments (Liess et al. 2006; Foit et al. 2012), no additional food was supplied to the systems after the end of the colonization period. Instead, nutrition came from periphyton growth on the sediments and suspended algae and bacteria in the water column. This set-up was chosen to mimic density dependent processes in natural systems (Halbach 1970) and enforce competition between the two test species. Only adult mosquitoes were provided with a saturated glucose solution to enable reproduction.

**Water Quality During the Pre-Exposure Period**

After 5 months of colonization, population monitoring of *C. pipiens* (once per week) and *D. magna* (twice per week) began. Those systems, with low emergence rates of adult *C. pipiens* were still populated with larvae and eggs to simulate spawning events for another two months. Physico-chemical parameters were very homogeneous across all systems in the pre-exposure period (temperature 20.2 ± 0.3 °C, conductivity 987 ± 84 µS/cm, oxygen 10.1 ± 0.5 mg/L, pH 7.3 ± 0.4). Nutrient levels were similar to previously conducted studies ($PO_4$: 0.2 ± 0.1 mg/L, $NO_3$: 0.7 ± 0.4 mg/L, $NO_2$: 0.02 ± 0.01 mg/L, $NH_4$: 0.03 ± 0.04 mg/L). The median suspended biomass of 0.2 mg/L was in the range of oligotrophic lakes, suggesting that most systems were strongly limited in biomass available for feeding, however measurements had a considerable range (90%-quantile: 0.01–2.97 mg/L). The environmental parameters that characterized the systems are summarized in Table S1 for the pre-exposure period and in Table 1 for the post-exposure period.

**Exposure**

After the 2-month pre-exposure period, the systems were exposed two times to the pyrethroid insecticide esfenvalerate with a recovery period of 1 month between exposures. The treatments consisted of 5 esfenvalerate exposure levels (solvent control, 0.1, 1, 10, 100 ng/L) with a treatment size of 16 replicates each. For the preparation of stock solutions, 5 mg esfenvalerate (CAS 66230-04-4, HPC Standards GmbH, Cunnersdorf, Germany) were dissolved in DMSO and diluted to a concentration of 1000 µg/L, which also served as exposure solution for the highest exposure treatment. From this stock, exposure solutions were diluted to 100µg/L, 10µg/L and 1 µg/L. An additional solution containing DMSO was prepared to serve as the exposure solution for the solvent control. The stocks were prepared on the day preceding the exposure and were kept refrigerated overnight. On the day of exposure, 350 µL of the treatment specific exposure solutions were added to the corresponding systems containing 3.5 L ADaM medium, amounting to a solvent concentration of 0.01 % v/v.

The accuracy of the exposure concentrations was determined by measuring the concentration of the stock solutions spiked to 1 L samples of freshly prepared ADaM. In addition, 50 ml water samples from 4 randomly selected replicates of the highest exposure treatment (100 ng/L) were taken exactly 1h after exposure and 48 h after exposure. Chemical analysis of the tested samples was performed by SGS Analytics Germany GmbH, using a GC-MS. Measured concentrations of stock solutions and samples of experimental replicates are shown in Table S3 and Table S4. The measured concentrations in experimental replicates in the highest esfenvalerate treatment are very homogeneously at 38.8 ± 11 ng/L, 1 h after exposure and are always below the limit of quantification (LOQ) of 20 ng/L after 48h. Rapid dissipation of esfenvalerate from the water column due to adsorption and photo degradation can explain the repaid decay in the first 48h hours after exposure.

**Biological assessment of Exposure Concentrations**

In addition to chemical analysis of the exposure solutions, the effect of esfenvalerate on standard test organisms was assessed. This was done under standard conditions and in the nanocosm medium. These standardized experiments were conducted in parallel to the exposure of the main experiment. For each test system, 2 x 25 ml beakers were filled with 20 ml samples of the test systems 1 hour after exposure. 5 neonates (< 24h) of *D. magna* were placed in one beaker and 5 larvae (< 96 h) of *C. pipiens* in the other and survival was observed for 48 h. During this period, organisms were not fed to approach conditions in the test systems. In addition, the same setup was prepared for each exposure concentration, plus a test concentration of 1000 ng/L, in standard ADaM medium. Populations were monitored for survival for 2 days without feeding (according to OECD acute standard test (OECD 2004)). Figure 1a shows that *C. pipiens* are slightly more sensitive (*Culex* $EC_{50}$ = 71 ng/L, *Daphnia* $EC_{50}$ = 176 ng/L) under standard conditions. When tested in the nanocosm medium, *Culex* had 10% higher control mortality also under non-lethal esfenvalerate concentrations (Fig.1b), while no control mortality was detected in *D. magna*.

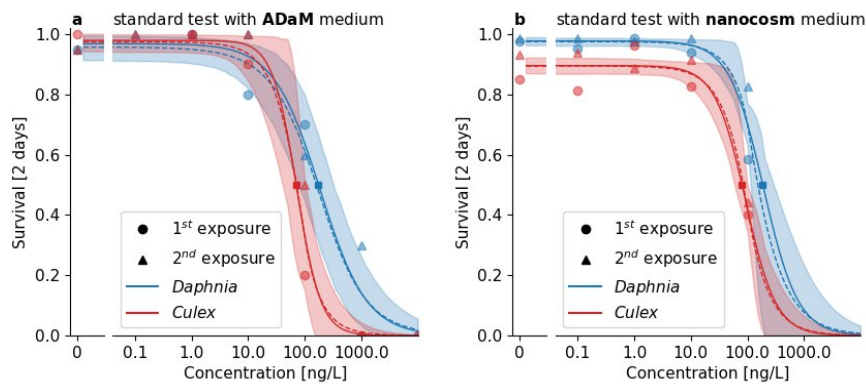

**Figure 1**. Similar sensitivity of *D. magna* (age < 24h) and *C. pipiens* (age < 96 h) after 48h of exposure to esfenvalerate. The reported esfenvalerate $EC_{50}$ for *D. magna* from EPA database is 310 ± 330 ng/L (Table S2). Survival data were obtained from standard tests conducted in parallel to the 1st and 2nd exposure of the nanocosm test systems with the same exposure solutions. **(a)** under standard conditions (*Culex* $EC_{50}$ = 71 ng/L, *Daphnia* $EC_{50}$ = 176 ng/L). **(b)** in samples of experimental units (nanocosms) taken 1h after exposure to esfenvalerate (*Culex* $EC_{50}$ = 80 ng/L, *Daphnia* $EC_{50}$ = 187 ng/L). The squares indicate the $EC_{50}$ and shaded areas show the Bayesian credible interval of the estimate. The solid line is the maximum likelihood estimate of a 3-parameter log-logistic function and the dashed line is the Bayesian fit.

**Monitoring of species abundance**

We monitored population development of *D. magna* by taking 3 images of each system with a Panasonic DC-FZ1000-II (Panasonic Corporation, Kadoma, Japan), twice per week with an improved image analysis technique compared to the approach developed by Foit et al. (2012).

First, motion was detected by background subtraction; this method is based on differences between two consecutive images. Background subtraction removed all static parts of the image, so that only objects moving in both images remained. Taking the element wise maximum of this difference resulted in only the moving objects of the first image. Depending on the amount of movement in the system, between 100–100 000 detection candidates were generated. Large numbers of proposals occurred when the background was even slightly moving, or the lighting conditions changed during capture. In a second step, the bounding boxes around the coordinates of the detection were analyzed for characteristic properties and stored in a file. These data points comprised the basis for the classification. 50 randomly selected images were annotated based on the candidate proposals. After annotation, a Support Vector Machine (SVM) classifier was trained with the annotated tags from the 50 images. The resulting accuracy of the unseen test set was 98%.

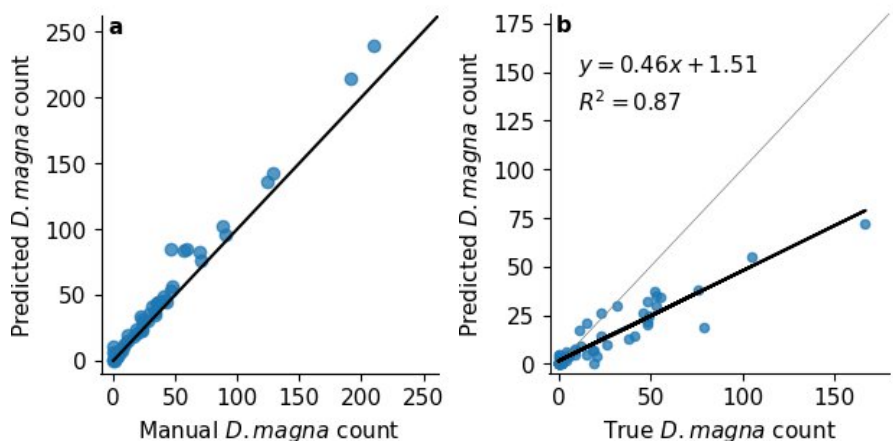

**Figure 2**. Validation of the detection method. The manual count indicates the number of organisms identified by visually counting *D. magna* in an image. In contrast, the true count is the actual number of organisms in the vessels determined when the experiment was ended. The predicted count is the number of organisms estimated by the classification algorithm. **(a)** Classifier evaluation of the capacity to detect organism from images. **(b)** Validation of the method by comparison of estimated organism count from image segmentation and classification with the true count from the last day of experimentation.

Of all labeled *Daphnia*, 97% were correctly classified as such, however an arbitrary detection candidate of a moving object generated a 2% chance of false positive detection, resulting in a slight tendency for over-detection (Fig. 2a). This resulted in problems if a large number of detection candidates was generated (e.g., when the camera was slightly moved during image capture). Therefore, in a 3$^{rd}$ step of the analysis, the image with lowest overall difference in pixels was chosen per series. This classification method was then used to detect population abundance in 7680 images taken throughout the experiment. Validation with the true organism count obtained at the end of the experiment shows that approximately 50% of the organisms are detected (Fig. 2b), however, this divergence is consistent throughout the assessed systems, which allows to assess the relative effects in the system.

The abundance of larvae of *C. pipiens* was manually counted once per week. Since larvae of *C. pipiens* generally remain static in their positions below the water surface, it was possible to determine accurate population counts. Also, in contrast to automatic detection methods it was easily possible to distinguish between exoskeletons of emerged larvae and their submerged siblings. In order to detect any organisms hiding in the sediments, the systems were gently moved to provoke escape reactions of *Culex* larvae. The abundance of C. larvae directly after hatching is not included in the population count. Only organisms > 1mm were included into the analyses.

Although, the fraction of *D. magna* in the water column was representative of the system state (Fig. 1b), future studies should use more homogeneous, dark sediments to facilitate the detection of organisms on the sediment. The automated detection of slow-moving organisms like *C. pipiens* could be enabled by using permanently installed cameras with longer intervals between images.

**Sampling and Measurement of Environmental Parameters**

Weekly, a 5.5 ml sample was taken to measure physico-chemical parameters and cell density. Every second week, additional 20 ml sample was taken to measure the nutrient status of the systems. Greatest care was taken to avoid cross contamination of the systems. Therefore, a syringe was connected with silicon tubing to each nanocosm and reused for the entire duration of the experiment. After sampling, the samples were stored cool until analysis at or measured directly and discarded thereafter. Sampling was conducted in parallel to the monitoring of the systems. Since this process took several hours, variations in the reported parameters due to daily temperature fluctuations are present in the dataset. Medium reductions due to sampling and evaporation were replenished with bi-distilled water. Major nutrient concentrations ($NO_2$, $NO_3$, $NH_4$, $PO_4$) were measured every second week with a Photometer (PF-12plus, Macherey-Nagel, Düren, Germany). To increase the accuracy, values were calculated from re-calibrated spectral absorption measurements and estimated concentrations (Fig. S1). Physico-chemical parameters were measured with a multi-parameter device (Portavo 908 Multi, Knick Elektronische Messgeräte GmbH & Co. KG, Berlin, Germany). Density of suspended cells were measured with a CASY-TTC cell counter (Schärfe Systems, Reutlingen, Germany, now OMNI Life Science, Bremen, Germany). The raw count data was passed through a filter, discarding measurements where total counts were < 2, to separate white noise from signal. Data were then smoothed and total volume in µL/L was calculated and estimated as suspended biomass density in mg/L assuming a wet weight density of 1 mg/µl (e.g. (Zhu et al. 2021)). The pre-exposure measurement values are reported in Table S1.

**Table 1**: average of environmental parameters per group during post-exposure. Values are reported as average across time and replicates and the associated standard deviation.

| variable | 0.0 ng/L | 0.1 ng/L | 1.0 ng/L | 10 ng/L | 100 ng/L |
|---|---|---|---|---|---|
| Temperature [°C] | 20.2 ± 0.3 | 20.1 ± 0.3 | 20.0 ± 0.3 | 20.0 ± 0.3 | 20.2 ± 0.3 |
| Conductivity [µS/cm] | 969 ± 49 | 977 ± 53 | 991 ± 53 | 997 ± 96 | 991 ± 78 |
| Oxygen saturation [mg/L] | 9.39 ± 0.27 | 9.43 ± 0.37 | 9.33 ± 0.3 | 9.38 ± 0.32 | 9.34 ± 0.32 |
| pH | 7.12 ± 0.08 | 7.11 ± 0.09 | 7.13 ± 0.09 | 7.1 ± 0.08 | 7.15 ± 0.16 |
| $PO_4$ [mg/L] | 0.45 ± 0.33 | 0.51 ± 0.39 | 0.77 ± 1.99 | 0.6 ± 0.89 | 0.65 ± 0.82 |
| $NO_3$ [mg/L] | 1.42 ± 0.69 | 1.47 ± 0.62 | 1.31 ± 0.62 | 1.44 ± 0.77 | 1.5 ± 0.83 |
| $NO_2$ [mg/L] | 0.03 ± 0.03 | 0.03 ± 0.04 | 0.04 ± 0.05 | 0.03 ± 0.05 | 0.05 ± 0.05 |
| $NH_4$ [mg/L] | 0.0 ± 0.01 | 0.01 ± 0.01 | 0.01 ± 0.02 | 0.01 ± 0.04 | 0.03 ± 0.15 |
| suspended biomass [mg/L] | 0.45 ± 1.0 | 0.41 ± 0.51 | 0.31 ± 0.3 | 0.67 ± 1.27 | 1.56 ± 5.8 |

The measured levels of N and P were in the range of eutrophic lakes (compare e.g. (Šorf et al. 2015; Beklioğlu et al. 2017)). However, continuously high levels of dissolved levels of oxygen and low densities of suspended cells indicate that the observed nutrient concentrations had no effect on the studied systems. Direct effects of these nutrients are also unlikely since, they were not near high enough to elicit direct effects on *D. magna* (Serra et al. 2019).

**Statistics**

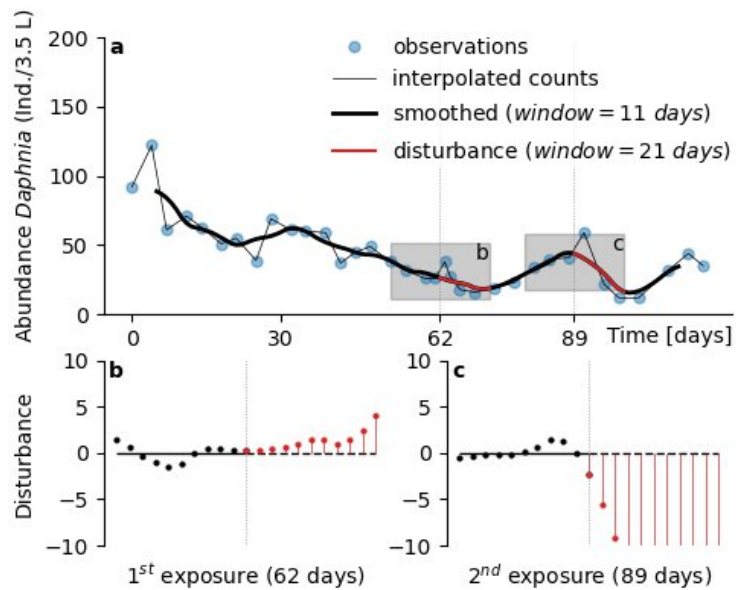

**Figure 3**: Time series smoothing and exemplary disturbance analysis of one experimental unit. **(a)** shows a running average (solid black line) that has been computed through the time series (blue dots). The lower panels show deviations from the linear pre-exposure trend that has been extrapolated (dashed line). **(b)** first exposure on the 3$^{rd}$ of June 2021 (day 62) **(c)** 2$^{nd}$ exposure on the 30$^{th}$ of June 2021 (day 89).

For the entire analysis 17 systems were excluded from subsequent analysis. The detailed reasons for removal are listed in the Supplementary Method S1. For all analyses considering the temporal dynamic of the systems, the time series were smoothed by computing centered running averages with a time window of 11 days (Fig. 3a). This was done to reduce the influence of very short termed fluctuations in the signal, which makes the analysis more robust to measurement errors, but may rarely underestimate true treatment effects such the saw tooth pattern visible in Figure 3a.

*Disturbance Analysis to Identify Short Term Effects of Esfenvalerate*

Due to the complexity of the analyzed systems, high variance between experimental replicates complicated the identification of general patterns in the time series. The following analysis was developed to robustly identify immediate effects of the tested esfenvalerate concentrations in replicated time series with high variance between replicates. In smoothed time series, 21–day long segments centered around the exposure events were isolated (Fig. 3a, gray boxes). Then linear trends were computed through the 10–day pre-exposure sections (Fig. 3b,c, solid horizontal lines) and were extrapolated to the following 11 days (Fig. 3b,c, dashed horizontal lines). Disturbances were then estimated by calculating differences between extrapolated pre-exposure trends and the true development of smoothed time series (Fig. 3b,c, red vertical lines).

The described analysis allows the estimation of low disturbances when the population development is characterized by smooth, non-volatile cycles, which we interpret as normal behavior. On the contrary, high variance in the signal will lead to strong disturbance signals and be indicative of treatment effects.

*Correlation Analysis to Identify Changes in Interspecific Competition*

In the study of interaction between species, measuring the correlations between species can give insight into their relationship (Moran 1953; Ranta et al. 1995; McCarthy 2011). Strongly positive correlations between abundance will in theory emerge, when both species equally respond to low or high levels of resource availability. In essence, when they are coexisting with significant overlap of shared resources. On the other hand, if the exclusion of either one species occurs, strong negative correlations between species will be observed. Natural systems, repeatedly observed over time, will show correlation coefficients between these extremes. However, trends in either one of the directions are indicative of changes in the relationship between species and will be interpreted as such in this work.

Estimating the correlation between count data is a non-trivial task. The approximation with the Pearson correlation coefficient will underestimate negative correlations due to the constraint of count data to be non-negative. In addition, multivariate Poisson distributions have been previously restricted to positive correlations (Ghosh et al. 2021). Modern Bayesian inference frameworks (Salvatier et al. 2016) allow for flexible transformations of variables, which enabled us to approach the problem by modeling the rate parameters of Poisson distributions as exponentiated multivariate normal variables.

$$
\begin{align}
N_s &\sim Poisson(\lambda_s) \tag{1} \\
\lambda_s &= e^{\log(\lambda_s)} \tag{2} \\
\log(\lambda_s) &\sim MultivariateNormal(\text{mu} = \mu_s, \text{covariance} = cov) \tag{3} \\
cov &\sim LKJ(\eta = 1, \sigma_s) \tag{4} \\
\mu_s &\sim Cauchy(0, 1) \tag{5} \\
\sigma_s &\sim HalfCauchy(1) \tag{6}
\end{align}
$$

Weakly informative Cauchy distributions with heavy tails (Gelman et al. 2008; McElreath 2015) were used as priors for $\mu_s$ and $\sigma_s$, which describes the log species occurrence rates and their intrinsic deviation. An LKJ prior with uniform probability density over the correlation between the species ($\eta$ = 1) was used as an uninformed prior for the covariance structure of the multivariate Normal. A calculation example for 3 imaginary test systems: assume the numbers of organisms (*Culex*, *Daphnia*) in the respective systems were (5, 10), (10, 20) and (20, 40). A correlation coefficient close to 1 would be estimated, although with large HDIs representing the uncertainty, since only 3 samples are given. In an opposing example were (1, 20), (50, 2), (0,0) are observed, a correlation coefficient near –1 would be estimated, representing the observation that at most one species was dominant. An estimation example for a simulated dataset is given in Figure S3, which shows that the correlation coefficient can be estimated very well, even with only 12 samples, which is representative for this study. The 95% highest posterior density interval (HPDI) was computed to calculate credible intervals, which are considered to be the Bayesian analog to confidence intervals. However, in contrast to confidence intervals, a 95% credible interval includes the true parameter value with a 95% probability by definition.

Interspecific correlation coefficients, including Bayesian uncertainty estimates were recovered from the covariance matrix, which were estimated for the whole pre- and post-exposure datasets (results: Fig. 6) and for each day in the smoothed time series (see Fig. 3) to obtain trends in the interspecific correlation (results: Figs. 4, 7).

## Results

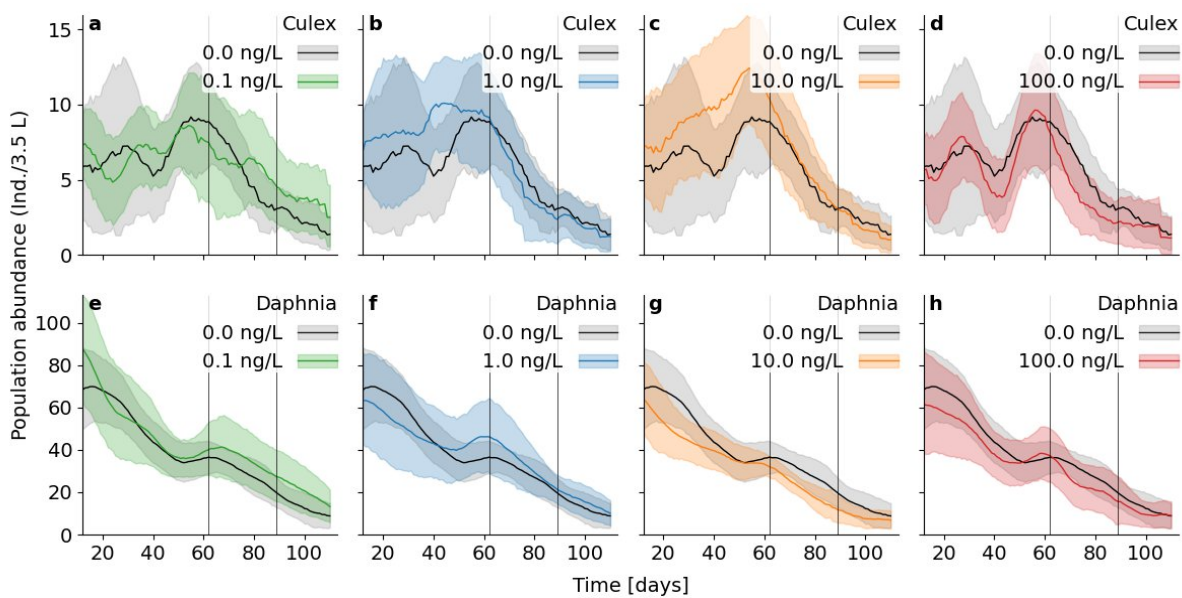

**Figure 4**. Species abundance per treatment over the time of the experiment in days computed with a Bayesian model of correlated, Poisson distributed variables. The vertical lines indicate the times of exposure to esfenvalerate. Shaded areas are 95% credible intervals and indicate the uncertainty of the estimates. **(a–d)** Expected abundance *C. pipiens*. **(e–h)** Expected abundance *D. magna.*

Figure 4 shows the development of population densities of the two competing species before and after exposure to esfenvalerate as a treatment average. Larvae populations of *C. pipiens* were stable or increasing in the pre-exposure period with highly variable population densities across replicates, indicated by large credible intervals (Fig. 4a–d). This is attributed to continued stocking of low density *Culex* populations with additional eggs and larvae in the pre-exposure period. Only when stocking was ceased in the post exposure period, negative trends were visible in the population density of *C. pipiens*. In contrast, *D. magna* populations, which were not artificially stocked in the pre-exposure period, follow a steady decline over the entire period of the experiment (Fig. 4e–h). In general, the declining population density reflect that the systems were characterized by resource scarcity. This corresponds to the low density of suspended organic matter (median 0.21 mg/L, 90%-quantile: 0.01–2.97 mg/L). As expected, due to the necessity of artificial stocking in the pre-exposure phase and slightly but significantly higher baseline mortality in nanocosm medium (Fig. 1b), *C. pipiens* were significantly less abundant than *D. magna* over the entire duration of the experiment.

After exposure to esfenvalerate, average trends of populations exposed to esfenvalerate did not significantly deviate from the controls. Correspondingly, the fraction of low-density populations towards the end of the experiment (≤10% of the pre-exposure maximum) did not differ between *C. pipiens* and *D. magna*. Also, the physico-chemical parameters were similar across all treatments in the post exposure period (Table 1). Compared to the pre-exposure period (Table S1), oxygen saturation slightly decreased by 7%, while the medium pH, conductivity and temperature did not change. Phosphate and Nitrate concentrations approximately doubled in the post-exposure period, while Nitrite and Ammonium did not change. Only suspended biomass differed among treatments; however, the differences are smaller than the standard deviations (Table 1).

**Community Response to Esfenvalerate Exposure**

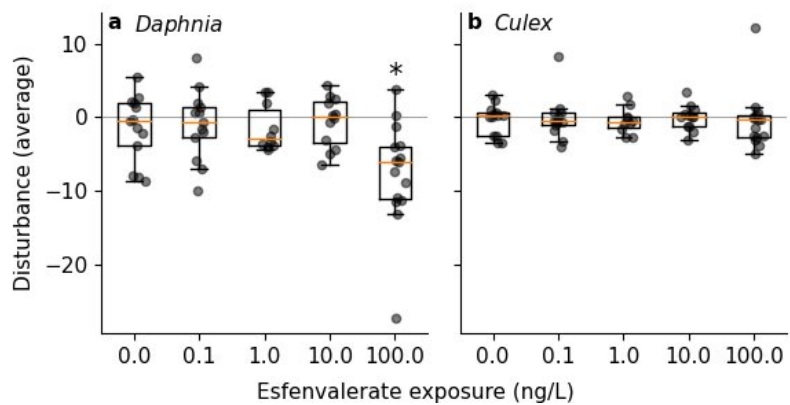

**Figure 5**. Average 11-day disturbance of competing populations calculated from the deviation of extrapolated pre-exposure trend (10 days) to observed post exposure development in a 21-day time window (see methods: disturbance analysis). **(a)** *Daphnia* disturbance after exposures to esfenvalerate. **(b)** *Culex* disturbance after exposure to esfenvalerate. A significant deviation from the control treatment is indicated by an asterisk.

Figure 5a shows that a concentration of 100 ng/L elicits a significant negative disturbance (-6.0, p = 0.02) on populations of *D. magna*. This is also visible in the volatile trajectory of Figure 4h (100 ng/L). While disturbances after exposure to 100ng/L were negative after both exposures, exposure to concentration ≤ 10 ng/L resulted in negative disturbances after the 1st exposure and positive disturbances after the 2nd exposure. On *C. pipiens*, exposure to esfenvalerate induced no significant short-term disturbances (Fig. 5b). Under standard conditions the species had similar sensitivities to esfenvalerate ($EC_{50}$ Culex = 80ng/L, $EC_{50}$ Daphnia = 180 ng/L, Fig. 1). However, these sensitivities were not reproduced on the community level, where *D. magna* is the only species significantly disturbed by exposure to 100 ng/L esfenvalerate. This could be explained by different durations of the observed post exposure period in standard tests (2 days) and the nanocosm test systems (11 days).

**Changes in Species Correlation after Exposure to Esfenvalerate**

It was a key question of this work, whether exposure to pesticide influences the interspecific competition at low concentrations. To answer this question, the correlation between abundance of both species over time was evaluated by applying Bayesian estimation of the covariance between two Poisson distributed variables.

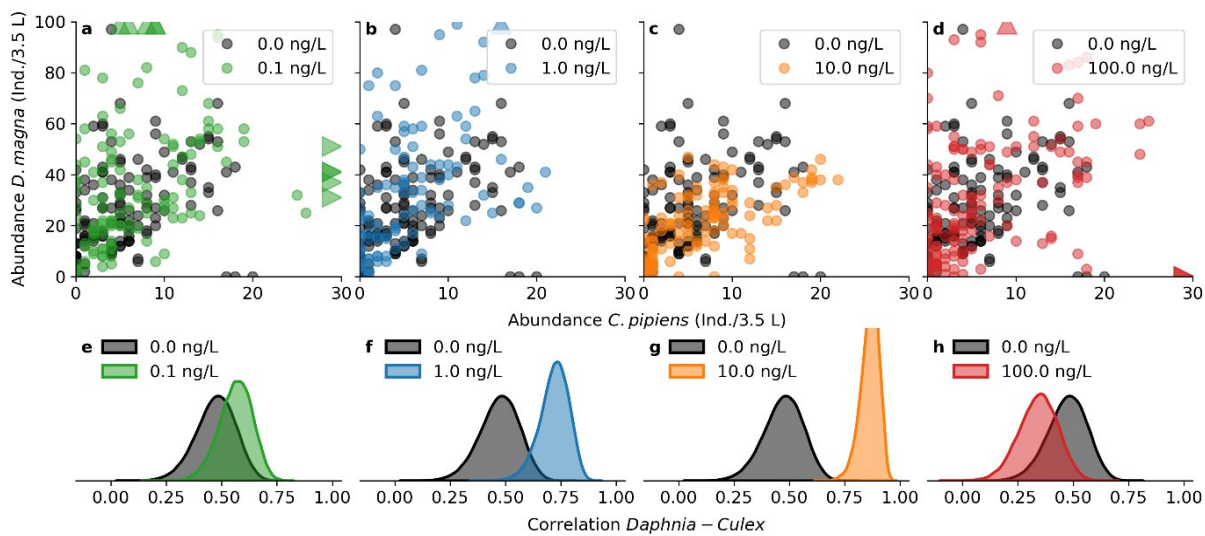

**Figure 6**. (**a–d**) Population densities of observed *Culex* and *Daphnia* during the entire post exposure period of all experimental replicates. The colored treatments are always compared to the same control dataset (gray). The displayed data-range was truncated to increase visibility of the dataset. Not shown data are indicated by triangles at the upper or right-hand side of the panels. (**e–h**) Bayesian posterior density estimates of the *Daphnia–Culex* correlation coefficient, fitted on the data in panels a–d with the model described in Equations 1–6. High correlations indicate that fluctuations in population

density were synchronized, while low correlations indicate that fluctuations in population density were not synchronized.

Figure 6a–d shows the pairs of population density of *C. pipiens* and *D. magna* at each observation in the post-exposure period. States with simultaneously high population densities of both *D. magna* and *C. pipiens* were rarely observed. Figure 6c shows the population densities of highly correlated species across multiple systems. The development of *C. pipiens* and *D. magna* populations in these replicates was synchronized, meaning that rarely one species was abundant while the other species was not. Figure 6e–h shows the estimated correlation coefficient between both species in the community. Small concentrations of esfenvalerate increased the correlation between competitors compared to the control treatment (Fig. 6e–g). This deviation is significantly positive over the entire post-exposure period in systems that were exposed to 10 ng/L (Fig. 6g). In contrast, exposure to 100 ng/L induced a slightly negative correlation shift between *Culex* and *Daphnia*. Considering the effect of 100 ng/L on the disturbance of *Daphnia* population (Fig. 5a), the reduction of correlation is a sign of extinction, also visible in the phase-space of Figure 6d, which shows that one or the other species become dominant, while the other is excluded.

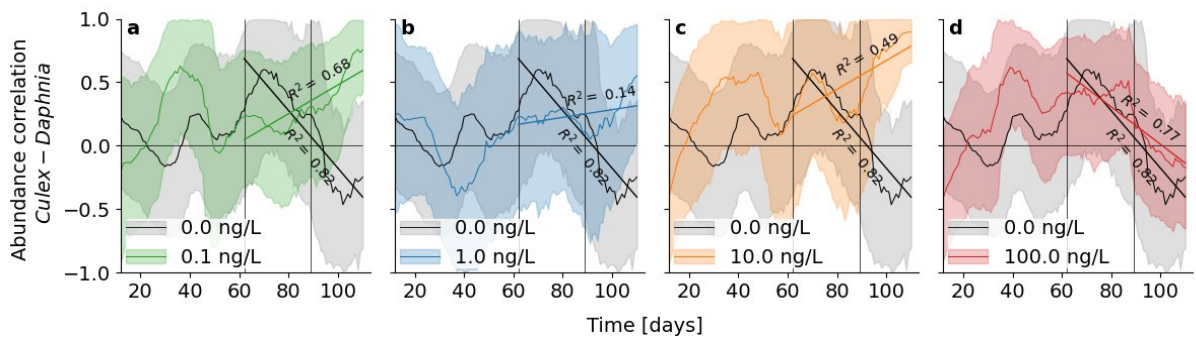

**Figure 7**. Temporal development of correlation coefficient between abundance of *C. pipiens* and *D. magna* (smoothed time series). Competitive exclusion in the control treatments and 100 ng treatments, and synchronized behavior in the three low concentrations. For each day of the time series, the correlation coefficient was estimated that best predicted the abundance pairs of *C. pipiens* and *D. magna* in all systems of one treatment. The shaded area shows the 95% credible interval and indicates the uncertainty of an estimate. The vertical lines indicate the times of exposure to esfenvalerate. Linear regression models were fitted to the correlations in the post-exposure period.

To identify the temporal development of interspecific competition, correlations between competitors were computed for each day in the monitoring period by fitting the model (Eqs. 1–6) on interpolated and smoothed daily observations. Linear regressions were computed to identify the treatment trends in competition in the post-exposure periods. We observed that the significantly negative trend in the control treatment emerged shortly after the exposure ($p < 0.001$), i.e., shortly after addition of manual stocking of *C. pipiens* larvae to the systems was stopped. Figure 7a–c shows that after exposure to 0.1–10 ng/L esfenvalerate correlations were significantly positive ($p ≤ 0.01$). However, the trend in the treatment exposed to 100 ng/L esfenvalerate was significantly negative ($p < 0.001$), although the correlations substantially dropped only after the second exposure (Fig. 7d).

**Effects of Environmental Conditions**

Neither physico-chemical parameters (e.g., temperature, oxygen) nor major nutrients varied among the treatments during the post-exposure phase (Table 1). While exposure to esfenvalerate significantly disturbed the population development of *D. magna*, the remaining unexplained variance remained large (Fig. 5a). Pre-exposure environmental parameters could not explain this variance (Fig. S2) as there were no significant correlations. Only the pH was mildly positively correlated with the disturbance residuals ($\rho = 0.27$). The concentration of major nutrients was the range of eutrophic lakes (Table S1), however no significant positive or negative correlation with the final abundance of *D. magna* or *C. pipiens* could be identified. Also, the correlation between pre-exposure environmental parameters and the final abundance of *D. magna* and *C. pipiens* was insignificant (Tables S5, S6).

**Discussion**

In this study we investigated the effect of environmentally realistic esfenvalerate exposures on a 2-species community in a highly competitive environment. We showed that exposing competing species with similar sensitivities to esfenvalerate results in a substantial reduction of interspecific competition at low concentrations, indicated by increasing positive correlations between species. These effects were detected far below effect concentrations established in standard tests, conducted in parallel to the experiment. The exposure to esfenvalerate increased the correlation between *D. magna* and *C. pipiens* with increasing levels of exposure, beginning as low as 3 orders of magnitude below the measured $EC_{50}$ (Fig. 6). Species correlations of treatments exposed to 0.1, 1 and 10 ng/L significantly increased over time during the post-exposure period. On the contrary, the concentration closest to the $EC_{50}$ (100 ng/L) decreased the correlation between species and also provoked significant disturbances in the population of *D. magna*.

The systems studied in this work were highly limited in suspended biomass available for nutrition (Tables 1, S1), due to the absence of external carbon inputs. For this reason, both species showed a similar declining population trend (Fig. 4). This decrease can also be attributed to low levels of primary production, approximated by the density of suspended biomass (Table S1). We assume that sufficiently large concentrations of N and P could not be converted to biomass in the studied systems. Possible reasons are strong competition of filter feeders, which prevented growth phases of phytoplankton, or insufficient lighting conditions. In the absence of suspended biomass, organisms were observed to graze on periphyton and biofilm, which were not quantified in this work but varied considerably among the experimental replicates. Resulting from this diversity, dominance and suppression of either species was approximately random, indicated by similar fractions of low density-populations, which led to negative correlations between the species' population densities. This is associated with high interspecific competition between *C. pipiens* and *D. magna* in the control treatments, which increased during the post-exposure period (Fig. 7) after colonization of C. pipiens was stopped. These results fit the theory that narrow environments with considerable niche overlap do not favor coexistence of competing species (Pastore et al. 2021).

**Exposure to high doses of esfenvalerate disturbs population and increases risk of single species dominance**

The exposure to esfenvalerate at 100 ng/L, induced significant, direct short-term disturbances in *Daphnia* populations and decreasing correlations in the post-exposure phase. Decreasing correlations indicate the suppression of one species, which could be exploited by the dominant species if the composition of the system in terms of suspended biomass, periphyton and biofilm allowed population growth. Since the variation between biomass density and other environmental parameters across experimental replicates could not explain the residual variance of the disturbance of species after exposure nor final population densities of either species, periphyton and biofilm may well have been responsible for the heterogeneity in the systems. Due to this heterogeneity of the systems, the observation of significant population level disturbance of esfenvalerate at 100 ng/L is assumed to be very robust. The direct effect of esfenvalerate at 100 ng/L is also visible in

Figure 6d, where species abundances increasingly converge to one or the other axes. Similar dynamic behavior has been observed for sub-populations of potato beetle larvae and adults (Costantino et al. 1997); when harvesting rates of adult beetles were experimentally increased, comparable to direct mortality effects of 100 ng/L esfenvalerate in the present work, populations were pushed out of equilibrium. Although the experimental conditions are only partly comparable, the results show that disturbances of competing (sub)populations can lead the way to significant changes in the dynamic of ecological communities.

**Exposure to low doses of esfenvalerate reduces interspecific competition**

From day 70, the control treatment showed a marked interspecific competition, indicated by decreasing correlations. In contrast, treatments exposed to low doses of esfenvalerate (0.1, 1, 10 ng/L) showed reduced interspecific competition, indicated by increasing correlations. This already occurred at concentrations more than 3 orders of magnitude below the $EC_{50}$ and reached its maximum at 10 ng/L (Fig. 6e-g, Fig. 7a-c). These observations support a recent simulation study, which predicts that the correlation between populations of different species increases if interspecific competition is absent because then development of both populations is driven only by environmental fluctuations (Lee et al. 2020). In contrast, the same study predicts decreases in interspecific correlations if high interspecific competition is present due to resulting suppression of one or the other species. Such mechanisms are precisely those observed in the present work. In another experimental study of marine environments, low pH led to altered competitive interactions between competing algae species and gradually led to a community shift (Kroeker et al. 2013). In a single species population study, exposure to 10 ng/L esfenvalerate reduced the competitiveness of *D. magna* and led to a hormetic increase in population abundance (Schunck and Liess 2023). Such an effect did not occur in this study. Instead, we assume that the presence of a competitor caused the absence of a stimulatory population effect, suggesting that findings of hormesis are dependent on the environmental context; i.e., only emerge when the environmental conditions do not penalize trade-offs associated with stimulatory effects.

**Conclusion and Outlook**

We show that concentrations 3 orders of magnitude below the $EC_{50}$ induced reductions in the interspecific competition between *D. magna* and *C. pipiens*. In contrast, concentrations near the $EC_{50}$ directly impacted *D. magna* populations and led to an increased tendency of single species dominance. The work also highlights, that single species sensitivity tests are insufficient to predict ecological effects on the community level. On the contrary, non-invasive population monitoring is very a promising approach, which can complement the higher tier risk assessment of ecological effects of toxicants, since the absence of sampling removes the most error prone and disturbing part of the method. By monitoring the correlation in abundance between competing species, more subtle effects can be detected and potentially hazardous long-term effects can be identified before they occur in the field.

## Supplementary Information

### CRediT authorship contribution statement

Florian Schunck: Conceptualization, Investigation, Data curation, Formal analysis, Visualization, Writing – Original draft. Matthias Liess: Conceptualization, Investigation – Guiding analytical cognition process, Writing.

### Acknowledgements

The authors thank Franz Dussl, Oliver Kaske, Maren Lück, Mavi Kotan and Naeem Shahid from the Department of System-Ecotoxicology of the Helmholtz Centre for Environmental Science for their support and advice during the conduct of the experiment. This work was partly funded from the European Union's Horizon Europe research and innovation programme under Grant Agreement No 101057014 (PARC).

## References

Alexander AC, Culp JM, Baird DJ, Cessna AJ. 2016. Nutrient-insecticide interactions decouple density-dependent predation pressure in aquatic insects. Freshwater Biology. 61(12):2090–2101. doi:10.1111/fwb.12711.

Allen J, Gross EM, Courcoul C, Bouletreau S, Compin A, Elger A, Ferriol J, Hilt S, Jassey VEJ, Laviale M, et al. 2021. Disentangling the direct and indirect effects of agricultural runoff on freshwater ecosystems subject to global warming: A microcosm study. Water Research. 190:116713. doi:10.1016/j.watres.2020.116713.

Arce-Funck J, Crenier C, Danger M, Cossu-Leguille C, Guérold F, Felten V. 2016. Stoichiometric constraints modulate impacts of silver contamination on stream detritivores: An experimental test with gammarus fossarum. Freshwater Biology. 61(12):2075–2089. doi:10.1111/fwb.12785.

Beklioğlu M, Bucak T, Coppens J, Bezirci G, Tavşanoğlu Ü, Çakıroğlu A, Levi E, Erdoğan Ş, Filiz N, Özkan K, et al. 2017. Restoration of eutrophic lakes with fluctuating water levels: A 20-year monitoring study of two inter-connected lakes. Water. 9(2):127. doi:10.3390/w9020127.

Costantino, Desharnais, Cushing, Dennis. 1997. Chaotic dynamics in an insect population. Science (New York, NY). 275(5298):389–391. doi:10.1126/science.275.5298.389.

Ebert D. 2022. Daphnia as a versatile model system in ecology and evolution. EvoDevo. 13(1):16. doi:10.1186/s13227-022-00199-0.

European Food Safety Authority. 2013. Guidance on tiered risk assessment for plant protection products for aquatic organisms in edge–of–field surface waters: EFSA panel on plant protection products and their residues (ppr). EFSA Journal. 11(7). doi:10.2903/j.efsa.2013.3290.

Fleeger JW, Carman KR, Nisbet RM. 2003. Indirect effects of contaminants in aquatic ecosystems. Science of The Total Environment. 317(1-3):207–233. doi:10.1016/S0048-9697(03)00141-4.

Foit K, Kaske O, Wahrendorf D-S, Duquesne S, Liess M. 2012. Automated nanocosm test system to assess the effects of stressors on two interacting populations. Aquatic toxicology (Amsterdam, Netherlands). 109:243–249. doi:10.1016/j.aquatox.2011.09.013.

Gause GF. 1936. The struggle for existence. Soil Science. 41(2):159. doi:10.1097/00010694-193602000-00018.

Gelman A, Jakulin A, Pittau MG, Su Y-S. 2008. A weakly informative default prior distribution for logistic and other regression models. The Annals of Applied Statistics. 2(4). doi:10.1214/08-AOAS191.

Gessner MO, Tlili A. 2016. Fostering integration of freshwater ecology with ecotoxicology. Freshwater Biology. 61(12):1991–2001. doi:10.1111/fwb.12852.

Ghosh I, Marques F, Chakraborty S. 2021. A new bivariate poisson distribution via conditional specification: Properties and applications. Journal of applied statistics. 48(16):3025–3047. doi:10.1080/02664763.2020.1793307.

Hardin G. 1960. The competitive exclusion principle. Science (New York, NY). 131(3409):1292–1297. doi:10.1126/science.131.3409.1292.

Hawlena H, Garrido M, Cohen C, Halle S, Cohen S. 2022. Bringing the mechanistic approach back to life: A systematic review of the experimental evidence for coexistence and four of its classical mechanisms. Frontiers in Ecology and Evolution. 10. doi:10.3389/fevo.2022.898074.

He F, Arora R, Mansour I. 2023. Multispecies assemblages and multiple stressors: Synthesizing the state of experimental research in freshwaters. WIREs Water.:1–19. doi:10.1002/wat2.1641.

Knillmann S, Stampfli NC, Beketov MA, Liess M. 2012. Intraspecific competition increases toxicant effects in outdoor pond microcosms. Ecotoxicology (London, England). 21(7):1857–1866. doi:10.1007/s10646-012-0919-y.

Knillmann S, Stampfli NC, Noskov YA, Beketov MA, Liess M. 2012. Interspecific competition delays recovery of daphnia spp. Populations from pesticide stress. Ecotoxicology (London, England). 21(4):1039–1049. doi:10.1007/s10646-012-0857-8.

Kroeker KJ, Micheli F, Gambi MC. 2013. Ocean acidification causes ecosystem shifts via altered competitive interactions. Nature Climate Change. 3(2):156–159. doi:10.1038/nclimate1680.

Lee AM, Saether B-E, Engen S. 2020. Spatial covariation of competing species in a fluctuating environment. Ecology. 101(1):e02901. doi:10.1002/ecy.2901.

Liess M, Foit K. 2010. Intraspecific competition delays recovery of population structure. Aquatic toxicology (Amsterdam, Netherlands). 97(1):15–22. doi:10.1016/j.aquatox.2009.11.018.

Liess M, Foit K, Becker A, Hassold E, Dolciotti I, Kattwinkel M, Duquesne S. 2013. Culmination of low-dose pesticide effects. Environmental science & technology. 47(15):8862–8868. doi:10.1021/es401346d.

Liess M, Liebmann L, Vormeier P, Weisner O, Altenburger R, Borchardt D, Brack W, Chatzinotas A, Escher B, Foit K, et al. 2021. Pesticides are the dominant stressors for vulnerable insects in lowland streams. Water Research. 201:117262. doi:10.1016/j.watres.2021.117262.

Liess M, Pieters BJ, Duquesne S. 2006. Long-term signal of population disturbance after pulse exposure to an insecticide: Rapid recovery of abundance, persistent alteration of structure. Environmental Toxicology and Chemistry. 25(5):1326–1331. doi:10.1897/05-466R.1.

MacArthur RH. 1958. Population ecology of some warblers of northeastern coniferous forests. Ecology. 39(4):599–619. doi:10.2307/1931600.

Mano H, Tanaka Y. 2016. Mechanisms of compensatory dynamics in zooplankton and maintenance of food chain efficiency under toxicant stress. Ecotoxicology (London, England). 25(2):399–411. doi:10.1007/s10646-015-1598-2.

McCarthy MA. 2011. Breathing some air into the single-species vacuum: Multi-species responses to environmental change. Journal of Animal Ecology. 80(1):1–3. doi:10.1111/j.1365-2656.2010.01782.x.

McElreath R. 2015. Statistical rethinking: A bayesian course with examples in r and stan.

Merritt RW, Dadd RH, Walker ED. 1992. Feeding behavior, natural food, and nutritional relationships of larval mosquitoes. Annual review of entomology. 37:349–376. doi:10.1146/annurev.en.37.010192.002025.

Moran PAP. 1953. The statistical analysis of the canadian lynx cycle. Australian Journal of Zoology. 1(3):291. doi:10.1071/ZO9530291.

OECD. 2004. Test no. 202: Daphnia sp. Acute immobilisation test. doi:10.1787/9789264069947-en.

Pastore AI, Barabás G, Bimler MD, Mayfield MM, Miller TE. 2021. The evolution of niche overlap and competitive differences. Nature ecology & evolution. 5(3):330–337. doi:10.1038/s41559-020-01383-y.

Pieters BJ, Liess M. 2006. Population developmental stage determines the recovery potential of daphnia magna populations after fenvalerate application. Environmental science & technology. 40(19):6157–6162. doi:10.1021/es052180o.

Ranta E, Lindstrom J, Linden H. 1995. Synchrony in tetraonid population dynamics. Journal of Animal Ecology. 64(6):767. doi:10.2307/5855.

Salvatier J, Wiecki TV, Fonnesbeck C. 2016. Probabilistic programming in python using pymc3. PeerJ Computer Science. 2:e55. doi:10.7717/peerj-cs.55.

Schunck F, Liess M. 2023. Ultra-low esfenvalerate concentrations increase biomass and may reduce competitiveness of daphnia magna populations. The Science of the total environment.:163916. doi:10.1016/j.scitotenv.2023.163916.

Serra T, Müller MF, Barcelona A, Salvadó V, Pous N, Colomer J. 2019. Optimal light conditions for daphnia filtration. The Science of the total environment. 686:151–157. doi:10.1016/j.scitotenv.2019.05.482.

Šorf M, Davidson TA, Brucet S, Menezes RF, Søndergaard M, Lauridsen TL, Landkildehus F, Liboriussen L, Jeppesen E. 2015. Zooplankton response to climate warming: A mesocosm experiment at contrasting temperatures and nutrient levels. Hydrobiologia. 742(1):185–203. doi:10.1007/s10750-014-1985-3.

Vaugeois M, Venturelli PA, Hummel SL, Accolla C, Forbes VE. 2020. Population context matters: Predicting the effects of metabolic stress mediated by food availability and predation with an agent- and energy budget-based model. Ecological Modelling. 416:108903. doi:10.1016/j.ecolmodel.2019.108903.

Zhu H, Liu XG, Cheng SP. 2021. Phytoplankton community structure and water quality assessment in an ecological restoration area of baiyangdian lake, china. International Journal of Environmental Science and Technology. 18(6):1529–1536. doi:10.1007/s13762-020-02907-6.